\documentclass[useAMS,usenatbib]{mnras}

\usepackage{threeparttable}
\usepackage{amsmath}
\usepackage{graphicx}
\usepackage{longtable}
\usepackage{amssymb}
\usepackage{tabularx, blindtext}
\usepackage{lscape}
\usepackage{url}
\newcommand{\teff}{\ensuremath{T_{\rm eff}}}

\title[EPIC247098361]{EPIC247098361b: a transiting warm Saturn on an eccentric $P=11.2$ days orbit around a $V=9.9$ star}
\author[R. Brahm et al.]{R.\ Brahm,$^{1,2}$\thanks{E-mail: rbrahm@astro.puc.cl}
                    N.\ Espinoza,$^{3,8}$ 
                    A.\ Jord\'an,$^{2,1,3}$
                    F.\ Rojas,$^{2}$ 
                    P.\ Sarkis,$^{3}$ 
 		            M.~R.\ D\'iaz,$^{4}$
		            \newauthor
                    M.\ Rabus,$^{2,3}$ 
                    H.\ Drass,$^{1,5}$ 
                    R.\ Lachaume,$^{2,3}$
                    M.~G.\ Soto,$^{4}$ 
                    J.~S.\ Jenkins,$^{4}$
                    M.~I.\ Jones,$^{6}$ 
                    \newauthor
                    Th.\ Henning,$^{3}$       
                    B.\ Pantoja,$^{4}$
                    M.\ Vu\v{c}kovi\'{c},$^{7}$
\\
$^{1}$Millennium Institute of Astrophysics, Santiago, Chile\\
$^{2}$Instituto de Astrof\'isica, Facultad de F\'isica, Pontificia Universidad Cat\'olica de Chile, Av. Vicu\~{n}a Mackenna 4860, 7820436 Macul,\\
Santiago, Chile\\
$^{3}$Max-Planck-Institut f\"{u}r Astronomie, K\"{o}nigstuhl 17, 69117 Heidelberg, Germany\\
$^{4}$Departamento de Astronom\'ia, Universidad de Chile, Camino El Observatorio 1515, Las Condes, Santiago, Chile\\
$^{5}$Center of Astro-Engineering UC, Pontificia Universidad Cat\'olica de Chile, Av. Vicu\~{n}a Mackenna 4860, 7820436 Macul, Santiago, Chile\\
$^{6}$European Southern Observatory, Casilla 19001, Santiago, Chile\\
$^{7}$Instituto de F\'isica y Astronom\'ia, Universidad de Vapara\'iso, Casilla 5030, Valpara\'iso, Chile\\
$^{8}$Bernoulli Fellow\\
}

\begin{document}

\date{Draft Version 4.0}

\pagerange{\pageref{firstpage}--\pageref{lastpage}} \pubyear{2002}

\maketitle

\label{firstpage}

\begin{abstract}
We report the discovery of EPIC247098361b using photometric data of the Kepler
K2 satellite coupled with ground-based spectroscopic observations. EPIC247098361b
has a mass of M$_{P}=0.397\pm 0.037$ M$_J$, a radius of R$_{P}=1.00 \pm 0.020$ R$_J$,
and a moderately low equilibrium temperature of $T_{eq}=1030 \pm 15$ K due to its
relatively large star-planet separation of $a=0.1036$ AU. EPIC247098361b orbits
its bright ($V=9.9$) late F-type host star in an eccentric orbit ($e=0.258 \pm 0.025$)
every 11.2 days, and is one of only four well characterized warm Jupiters having
hosts stars brighter than $V=10$.
We estimate a heavy element content of 20 $\pm$ 7 M$_{\oplus}$ for EPIC247098361b,
which is consistent with standard models of giant planet formation. The bright
host star of EPIC247098361b makes this system a well suited target for
 detailed follow-up observations that will aid in the study of the
atmospheres and orbital evolution of giant planets at moderate separations
from their host stars.

\end{abstract}


\section{Introduction}
\label{sec:intro}

Transiting hot Jupiters (giant planets with periods P$<$10d) have been efficiently
detected by several ground- and space-based surveys \citep[e.g.,][]{bakos:2004,pollacco:2006,
borucki:2010,bakos:2013}. This great number of discoveries has been key for constraining
theories of their formation, structure and evolution \citep[for a recent review see][]{dawson:2018}, but several unsolved 
theoretical challenges have emerged from these observations as well.
For example,  the specific source of the inflated radius of highly irradiated hot Jupiters
is a topic of active research. While several mechanisms have been proposed
\citep[for a review see][]{spiegel:2012},
their validation is not straightforward because in most cases the structural
composition (i.e. heavy element content) of these planets is not known, and therefore the
problem becomes degenerate.

Another long standing theoretical challenge is the actual existence of these
massive planets at short orbital separations, because most theoretical models of
formation require that Jovian planets are formed beyond the snow line where solid
embryos are efficiently accreted \citep{rafikov:2006}. While some orbital displacement
of the planet due to exchange of angular momentum with a gaseous proto-planetary disc
is expected to happen, it is not clear that this type of interaction can account for
the currently known population of giant planets with semi-major axes shorter than 1~AU.
One particular challenge is that a significant fraction of the hot Jupiter systems have been
found to have large spin-orbit angles, which are not expected to arise in a gentle disc
migration scenario \citep[for a review see ][]{winn:2015} .
While high eccentricity migration models predict the existence of highly misaligned
spin-orbit angles, a direct comparison between the model predictions and the obliquity distribution of hot Jupiters can be affected by the possible realignment of the
outer layers of the star due to tidal and/or magnetic interactions \citep{dawson:2014,li:2016}.

Transiting warm Jupiters (giant planets with periods $P>10$ d) are valuable systems
in the above mentioned context.
Due to their relatively long planet-star separations ($a \gtrsim 0.1$ AU), the internal structure
of warm Jupiters is not significantly affected by the tidal, magnetic and/or radiative
mechanisms that can significantly affect hot Jupiters. For this reason, theoretical models can be used
to investigate the internal composition of giant planets and how this depends on the global properties of the system (i.e., stellar mass, [Fe/H], multiplicity) in a more straightforward fashion.
Along the same line, given that for warm Jupiters the planet-star interaction is in general not
strong enough to realign the outer layers of the star, they are better suited systems
to test the predictions of high eccentricity migration models by studying the distribution
of spin-orbit angles \citep{petrovich:2016}.

Unfortunately, the detection of warm Jupiters around bright stars is hindered by
strong detection biases. The transit probability is proportional
to $a^{-1}$, and additionally the duty cycle required to discover transiting planets
with periods longer than 10 days is usually too high for typical ground-based
surveys, which are the ones that have made the most significant contribution to the
population of transiting giant planets with precisely determined masses and radii.
One workaround to this problem is to build longitudinal networks of identical telescopes 
to counteract the diurnal cycle \citep[e.g. HATSouth,][]{bakos:2013}. This configuration
has allowed the detection of planets with periods as long as 16 days \citep{brahm:2016:hs17}
Another solution is the use of space-based telescopes. Due to the precise and continuous $\approx$2 month observations per field, the Kepler K2 mission \citep{howell:2014}
is able to detect warm Jupiters \citep{smith:2017, shporer:2017}.
Additionally, it has an increased probability of detecting these type of system
on bright stars compared to the original \textit{Kepler} mission because it surveys a
larger area of the sky.

In this study we present the discovery of a warm Saturn around a bright star with the \textit{Kepler} telescope.
This discovery was performed in the context of the
K2CL collaboration, which uses spectroscopic facilities located in Chile to confirm and characterize transiting planets from K2 \citep{brahm:2016:k2,espinoza:2017:k2,jones:2017,soto:2018}. The structure of the paper is as follows. In \S~\ref{sec:obs} we present the
photometric and spectroscopic observations that allowed the discovery of EPIC247098361b,
in \S~\ref{sec:ana} we derive the planetary and stellar parameters, and we discuss our findings  in \S~\ref{sec:disc}. 

\section[]{Observations}
\label{sec:obs}

\subsection{Kepler K2}
\label{sec:k2}
\begin{figure*}
 \includegraphics[width=2\columnwidth]{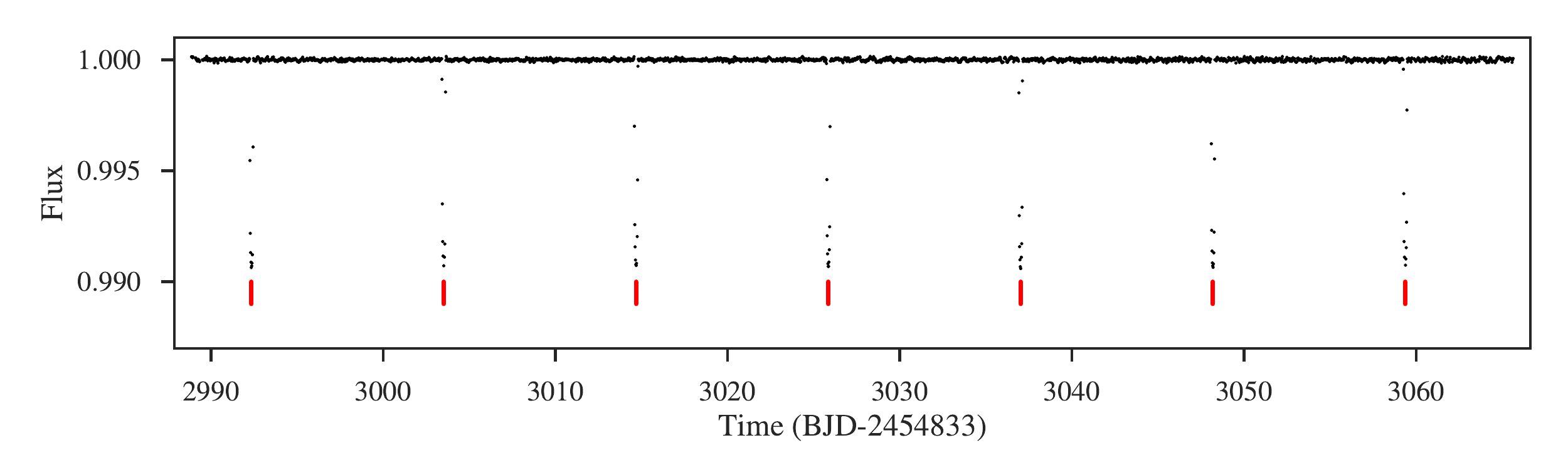}
 \caption{Detrended K2 photometry of EPIC 247098361. The transits of our planet candidate have been identified by red marks below each event.}
 \label{fig:photometry}
\end{figure*}

EPIC247098361 was observed by the Kepler K2 mission between March and May 2017, while carrying out the 
monitoring for campaign 13. The photometric data was reduced from pixel-level products using the EVEREST algorithm \citep{EVEREST1,EVEREST2}. Long-term trends in the data are corrected with a gaussian-process regression. Transiting planet detection is performed by using the
Box-fitting Least Squares algorithm \citep{BLS} on the processed light curves.
With this procedure we identified a $11.17$ day periodic signal, with a depth
consistent with that of a giant planet transiting a main sequence star.
The (detrended) K2 photometry for this target star is shown in Figure~\ref{fig:photometry}.
Due to the clear box-shaped transits and the brightness of the star, EPIC247098361
was selected as a high priority target for spectroscopic follow-up observations.

\subsection{Spectroscopic Observation}

\label{sec:spec}

High resolution spectroscopic observations are required to characterize the
host star, identify possible false positive scenarios, and to confirm the
planetary nature of the transiting companion via mass determination from the
radial velocity signal. The spectroscopic facilities that were used in this
work are summarized in Table~1, along with the main
properties of the observations.

\begin{table*}
 \centering
 \begin{minipage}{180mm}
  \caption{Summary of spectroscopic observations for EPIC247098361.}
  \begin{tabular}{@{}ccccccc@{}}
  \hline
   Instrument     &   UT Date(s)  & N Spec. &   Resolution &  S$/$N range & $\gamma_{RV} $[km s$^{-1}$] &  RV Precision [m s$^{-1}$] \\
 \hline
 Coralie / 1.2m Euler/Swiss   & 2017 Oct 31 -- Nov 2 & 3 &60000 & 27 -- 44 & 22.327 & 10 \\
 FEROS / 2.2m MPG            & 2017 Oct 03 -- 2018 Jan 28 & 18 & 50000 & 106 -- 167 & 22.398 & 7 \\
 HARPS / 3.6m ESO             & 2017 Nov 01 -- 2017 Nov 08 & 8 &  115000 & 36 -- 50 & 22.416 & 6 \\
\hline
\end{tabular}
\end{minipage}
 \label{tab:specobs}
\end{table*}

We obtained three spectra of EPIC247098361 with the Coralie spectrograph \citep{CORALIE} mounted
on the 1.2m Euler/Swiss telescope located at the ESO La Silla observatory.
Observations were obtained on three consecutive nights in October 2017, and 
they were acquired with the simultaneous calibration mode \citep{baranne:96}
where a secondary fiber is illuminated by a Fabry-Perot etalon in order to
trace the instrumental velocity drift produced by the changes in the environmental 
conditions of the instrument enclosure.
Coralie data was processed and analyzed with the CERES automated package \citep{jordan:2014,brahm:2017:ceres}. On top of the reduction and optimal
extraction of the spectra, CERES delivers precision radial velocity and
bisector span measurements by using the cross-correlation technique, and
an initial estimate of the atmospheric parameters by comparing the reduced
spectra with a grid of synthetic models \citep{coelho:2005}.
These three spectra allowed us to conclude that EPIC247098361 is a dwarf star 
($\log(g)\approx 4.2$) with an effective temperature of T$_{eff}\approx 5900$ K.
Additionally, there was no evidence of additional stellar components in the
spectra that could be linked to blended eclipsing binary scenarios, and the
radial velocity measurements rejected the presence of large velocity variations
caused by a stellar mass orbital companion. These properties boosted the follow-up observations of EPIC247098361 and we proceeded to obtain spectra with more powerful facilities.

\begin{figure*}
 \includegraphics[width=\textwidth]{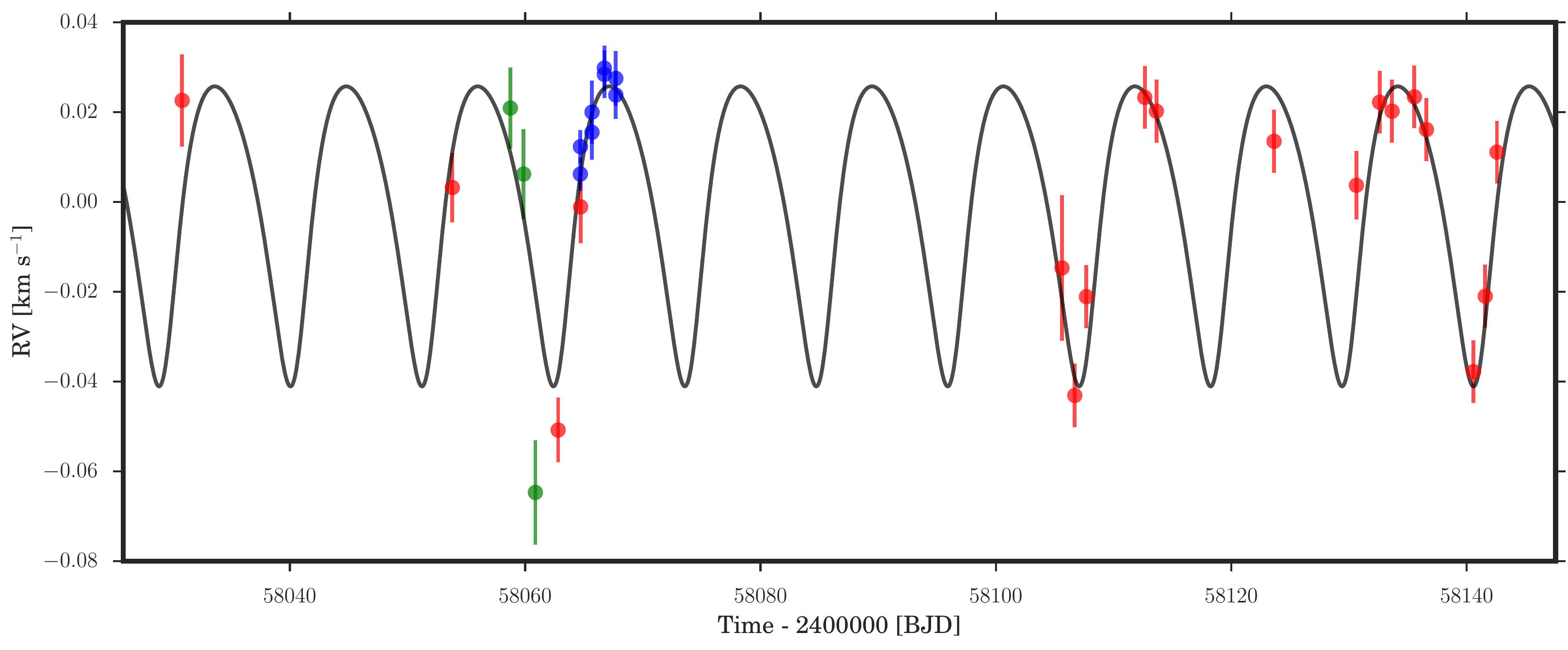}
 \caption{ Radial velocity (RV) curve obtained with FEROS (red), Coralie (green) and HARPS (blue). The black line corresponds to the Keplerian model with the posterior parameters found in Section 3. }
 \label{rvs-t}
\end{figure*}

\begin{figure}
 \includegraphics[width=\columnwidth]{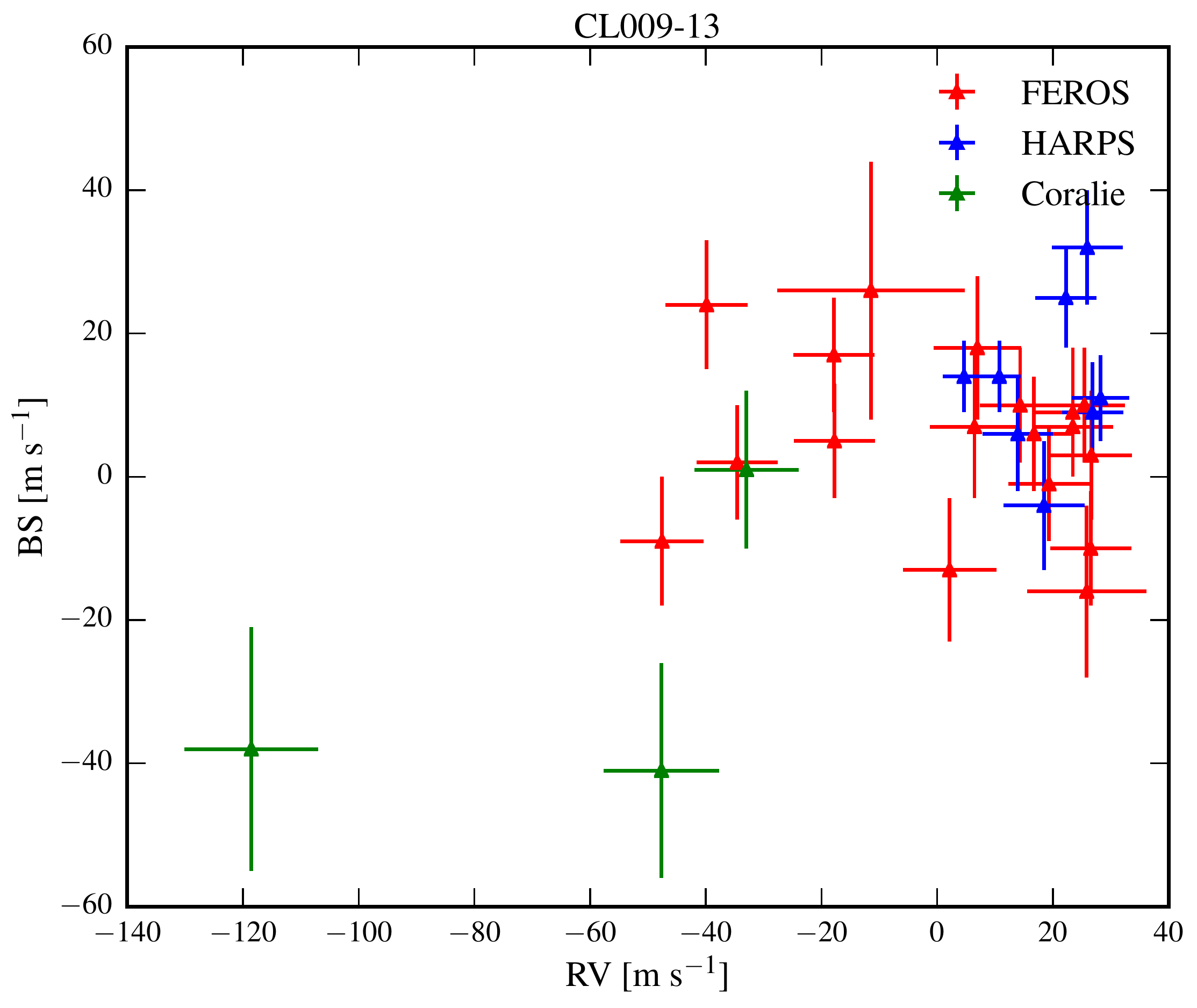}
 \caption{ Radial velocity (RV) versus bisector span (BIS) scatter plot using data from our spectroscopic observations of EPIC247098361. No significant correlation was found. }
 \label{rvs-bs}
\end{figure}

We obtained 18 spectra of EPIC247098361 between October of 2017 and January 2018
with the FEROS spectrograph \citep{kaufer:99} mounted on the MPG2.2m telescope,
and another eight spectra of the same target in November 2017 with the HARPS
spectrograph \citep{mayor:2003} mounted on the ESO 3.6m telescope. Both facilities
are located at the ESO La Silla Observatory. The FEROS observations were
performed with the simultaneous calibration mode where a Thorium-Argon lamp
illuminates a second fiber during the science observations. 
Given that the nightly instrumental drift of the HARPS spectrograph is
significantly smaller than the expected radial velocity variation produced
by a giant planet, the secondary fiber of this instrument was not used to
trace the velocity drift. Reductions and analysis of FEROS and HARPS spectra
were performed with the CERES automated package. The radial velocity
and bisector span measurements are presented in Table~4, and the
radial velocity curve is plotted in Figure~\ref{rvs-t}
As can be seen in this figure, the velocities obtained with the three
instruments are consistent with the radial velocity variation produced by a
giant planet with an eccentric orbit.
Additionally, no significant correlation was detected between the radial
velocities and bisector span measurements, as can be seen in
the radial velocity vs. bisector span (BIS) scatter plot on Figure~\ref{rvs-bs}.
We computed the distribution for the Pearson correlation coefficient between
the radial velocities and bisector span measurements, finding that it lies between
-0.13 and 0.65 at 95\% confidence level, and is therefore consistent with
no correlation. These spectroscopic observations allowed us to confirm that
the transit-like signal observed in the K2 data is produced by a planetary
mass companion.

\section{Analysis}
\label{sec:ana}
\subsection{Stellar parameters}
\label{sec:stellar-parameters}
We used the co-added FEROS spectra to estimate the stellar atmospheric parameters of EPIC247098361 by 
using the ZASPE code \citep{brahm:2015,brahm:2016:zaspe}. ZASPE determines T$_eff$, $\log(g)$, [Fe/H], and vsin(i) by
comparing the observed spectra to synthetic ones in the spectral regions most sensitive to changes in those parameters.
Additionally, reliable uncertainties are obtained from the data by performing Monte Carlo simulations that take into account the systematic mismatches between data and models.
Using this procedure we obtain the following parameters:
 T$_{eff}$ = 6020 $\pm$ 83 K, $\log(g)$ = 4.22 dex, [Fe/H] = 0.04 dex, and vsin(i) = 4.0 km s$^{-1}$, which are consistent with the initial estimates provided by CERES.
 
\begin{figure*}
 \includegraphics[width=\textwidth]{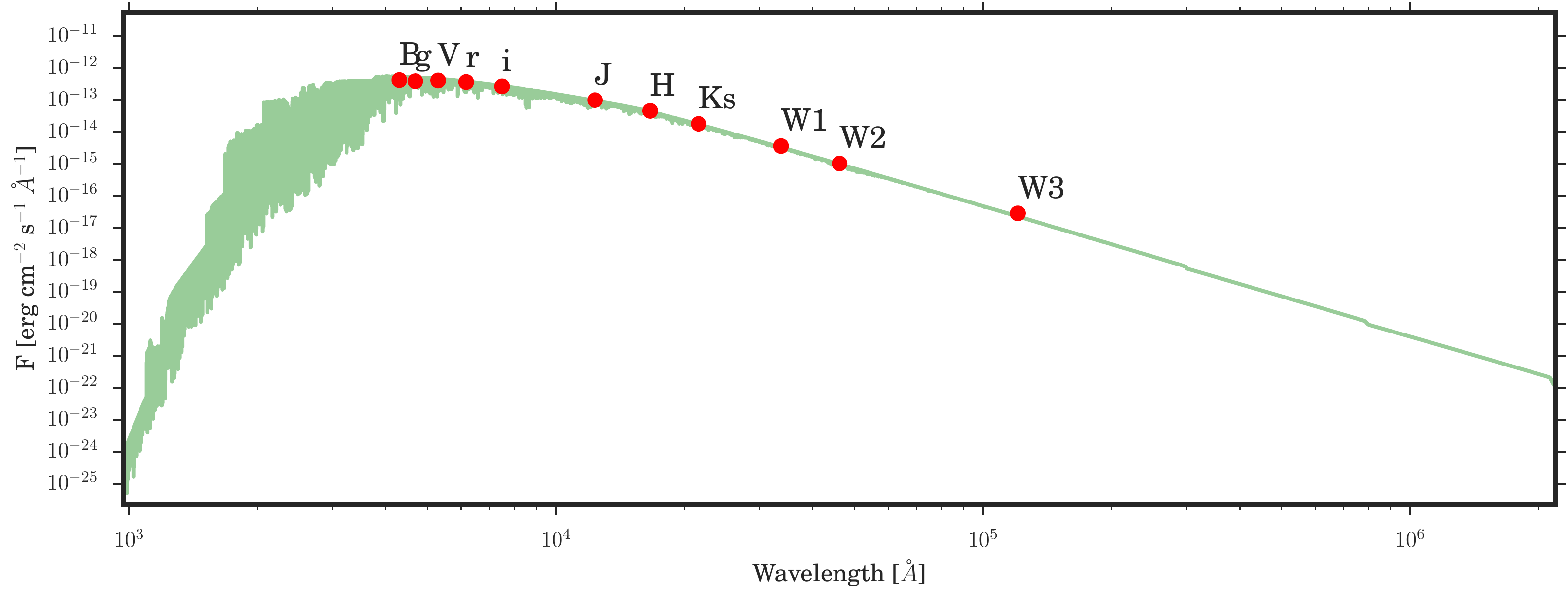}
 \caption{Spectral energy distribution of the \texttt{BT-Settl-CIFIST} model with atmospheric parameters similar to EPIC247098361. The observed
 flux densities for the different passband filters are identified as red circles.}
 \label{sed}
\end{figure*}

EPIC247098361 was observed by GAIA and its parallax is given on its DR1 \citep[$p$ = 7.69 $\pm$ 0.27 $mas$,][]{gaia:2016, gaia:2016:dr1}. We used this parallax value coupled to the
reported magnitudes in different bandpass to estimate the stellar radius, following an approach similar  to \citet{barragan:2017}. 
Specifically, we used the \texttt{BT-Settl-CIFIST} spectral models from \citet{baraffe:2015},   interpolated in \teff and log(g),
to generate a synthetic spectral energy distribution (SED) consistent with the atmospheric parameters of EPIC247098361.
We then integrated the SED in different spectral regions to generate synthetic magnitudes that were weighted by the corresponding
transmission functions of the passband filters. 
The synthetic SED along with the observed flux density  in the different filters are plotted in Figure \ref{sed}.

\begin{figure}
 \includegraphics[width=\columnwidth]{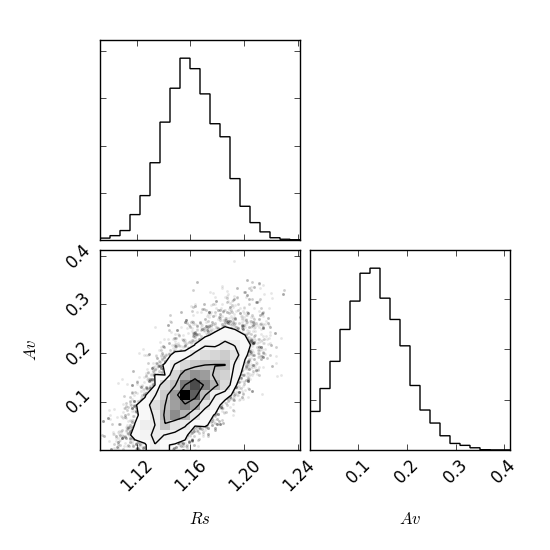}
 \caption{Triangle plot for the posterior distributions of R$_{\star}$ and $A_V$ obtained from the observed magnitudes and GAIA parallax.}
 \label{rstar}
\end{figure}

\begin{figure}
 \includegraphics[width=\columnwidth]{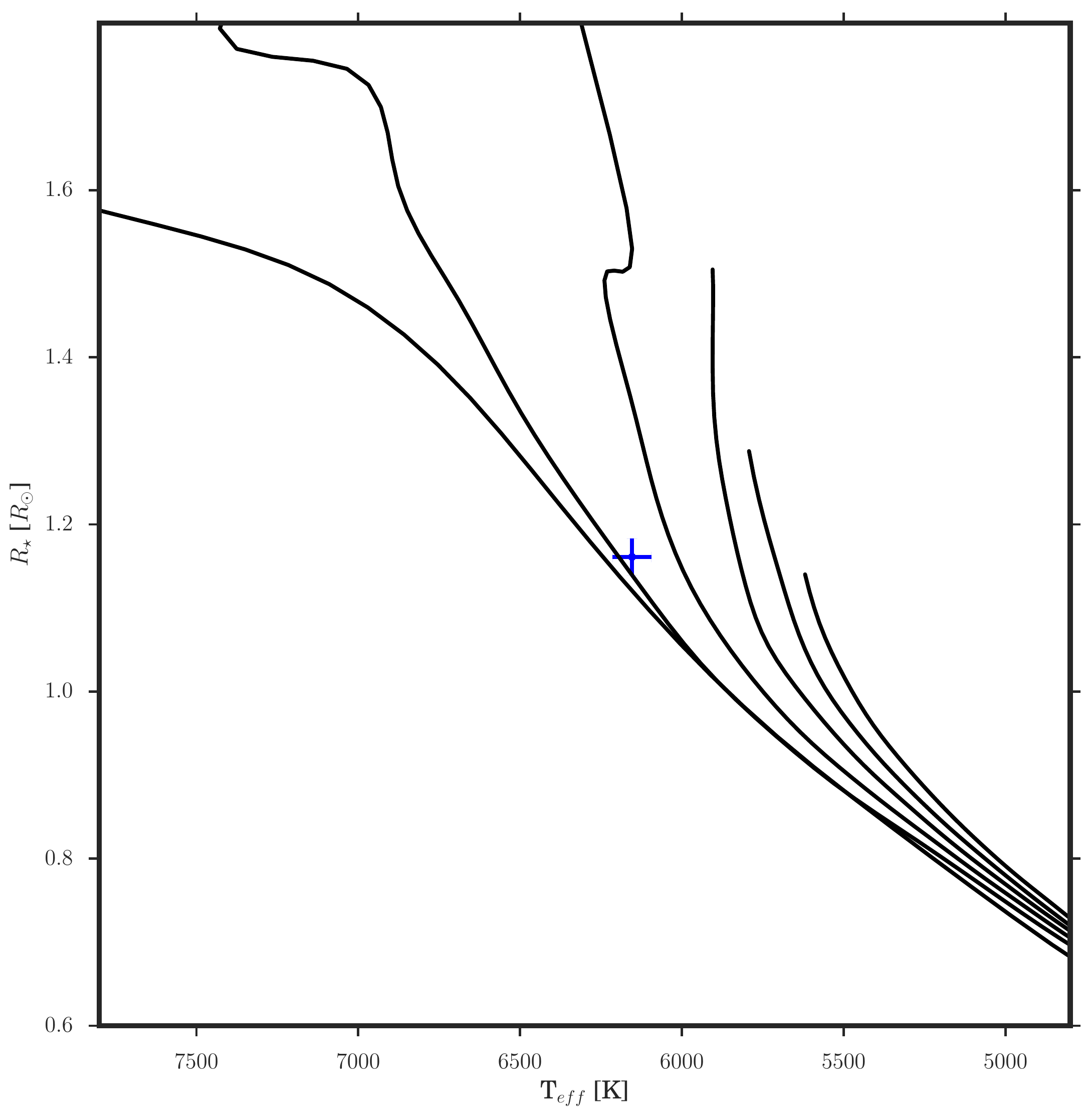}
 \caption{Yonsei-Yale isochrones for the metallicity of EPIC247098361 in the T$_{eff}$--R$_{\star}$ plane. From left to right the isochrones corresponds to 0.1, 1, 3, 5, 7, 9 Gyr.
 The position of EPIC247098361 in this plane is shown with a blue cross.}
 \label{iso}
\end{figure}

These synthetic magnitudes were used to infer the stellar radius (R$_{\star}$) and the extinction factor (A$_V$) by comparing them to the observed
magnitudes after applying a correction of the dilution of the stellar flux due to the distance. Specifically, our data was the stellar luminosity obtained
by multiplying the observed flux density $F_{\rm obs}^{\lambda_i}$  at the different passband filters ($\lambda_i$) with the square of the distance inferred from the GAIA parallax ($D$):
\begin{equation}
L_{\rm obs}  = 4 \pi F_{\rm obs}^{\lambda_i}  D^{2} .
\end{equation}
While the adopted model was:
\begin{equation}
L_{\rm mod}= 4 \pi F_{\rm syn}^{\lambda_i} R_{\star}^{2} e^{-A(\lambda_i)/2.5},
\end{equation}
where  $F_{\rm syn}^{\lambda_i}$ is the synthetic flux density at the different passband filters, and $A(\lambda_i)$ is the 
wavelength dependent extinction factor, which we take to be a function of visual extinction ($A_V$) as described in  \citet{cardelli:89}.
We used the \texttt{emcee} \texttt{Python} package \citep{emcee:2013} to sample the posterior distribution of  R$_{\star}$ and $A_V$.
Figure~\ref{rstar} shows the posterior distribution for the parameters. The estimated stellar radius from the parallax measurement was coupled to the  
estimated T$_{eff}$ to obtain the mass and evolutionary stage of the star by using the Yonsei-Yale isochrones \citep{yi:2001}. Figure~\ref{iso} shows
the isochrones in the  T$_{eff}$--R$_{\star}$ plane for different ages, with the values for EPIC247098361 indicated with a blue cross. This analysis allowed us
to obtain a more precise estimate for the stellar $\log(g)$ than the value obtained with ZASPE. This new $\log(g)$ value was held fixed in a new ZASPE iteration,
which was followed by a new estimate of the stellar radius and a new comparison with the theoretical isochrones. After this final iteration,
the stellar $\log(g)$ value converged to 4.389 $\pm$ 0.017, and the other stellar properties to the values listed in Table~2.
We found that EPIC247098361 is a late F-dwarf star (M$_{\star}$=1.192 $\pm$ 0.025 M$_{\odot}$, R$_{\star}$=1.161 $\pm$ 0.022 R$_{\odot}$)
and that it is slightly metal rich ([Fe/H]=0.1 $\pm$ 0.04).

\begin{table}
\label{tab:stellar}
 \centering
  \caption{Stellar properties and parameters for EPIC247098361.}
  \begin{tabular}{@{}lcc@{}}
  \hline
   
  Parameter      &   Value &  Method / Source \\
 \hline
 \\
Names     & EPIC247098361 & -- \\
RA            & 04:55:03.96 & -- \\
DEC         & 18:39:16.33 &   \\
Parallax  [$mas$]    & 7.69 $\pm$ 0.27 & GAIA\\
\hline
\\
$K_p$ (mag) & 9.789 & EPIC\\
B (mag) &10.469 $\pm$ 0.029 & APASS\\
g (mag) &10.286 $\pm$  0.184 & APASS\\
V (mag) & 9.899 $\pm$  0.039 & APASS\\
r (mag) & 9.749 $\pm$  0.033 & APASS\\
i (mag) & 9.663 $\pm$  0.011  & APASS \\
J (mag) & 8.739 $\pm$ 0.025 & 2MASS\\
H (mag) & 8.480 $\pm$ 0.011 & 2MASS\\
Ks (mag) & 8.434 $\pm$ 0.014& 2MASS\\
W1 (mag) & 8.380 $\pm$0.024& WISE\\
W2 (mag) & 8.419 $\pm$ 0.019& WISE\\
W3 (mag) & 8.391 $\pm$ 0.027& WISE\\
\hline
\\
T$_{eff}$  [K] &  6154 $\pm$ 60 & ZASPE \\
log(g) [dex]     &  4.379 $\pm$ 0.017 & ZASPE \\
$[$Fe/H] [dex]    &  0.10 $\pm$ 0.04 & ZASPE \\
$v \sin{i}$  [km s$^{-1}$] & 4.16 $\pm$ 0.282 & ZASPE \\
\hline
\\
M$_\star$ [M$_\odot$] &  1.192$_{-0.024}^{+0.025}$ & ZASPE + GAIA + YY \\
R$_\star$ [R$_\odot$] & 1.161$_{-0.021}^{+0.023}$ & ZASPE + GAIA  \\
L$_\star$  [L$_\odot$] & 1.718$_{-0.086}^{+0.101}$ & ZASPE + GAIA  + YY\\
Age    [Gyr]        &    1.26$_{-0.74}^{+0.71}$ & ZASPE + GAIA + YY \\
A$_V$  &  0.129$_{-0.062}^{+0.065}$ & ZASPE + GAIA  \\
\hline

\end{tabular}
\end{table}

\subsection{Global modelling}

\begin{figure}
 \includegraphics[width=\columnwidth]{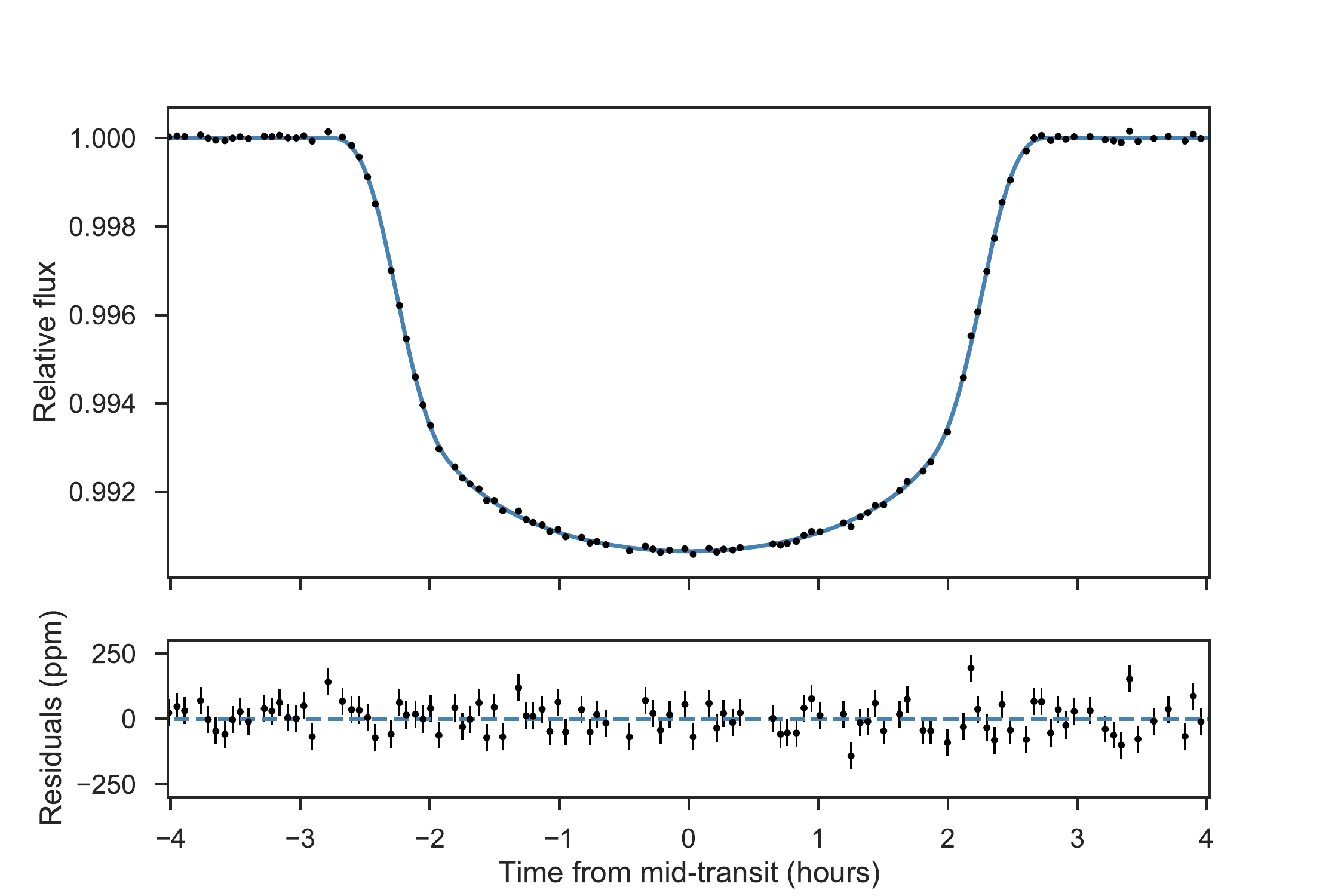}
 \caption{ The top panel shows the phase folded Kepler $K2$ photometry (black points) as a function of time at the time of transit for EPIC247098361b,
 and the model constructed with the derived parameters of exonailer (blue line). The bottom panel shows the corresponding residuals.}
 \label{exonailerlc}
\end{figure}

\begin{figure}
 \includegraphics[width=\columnwidth]{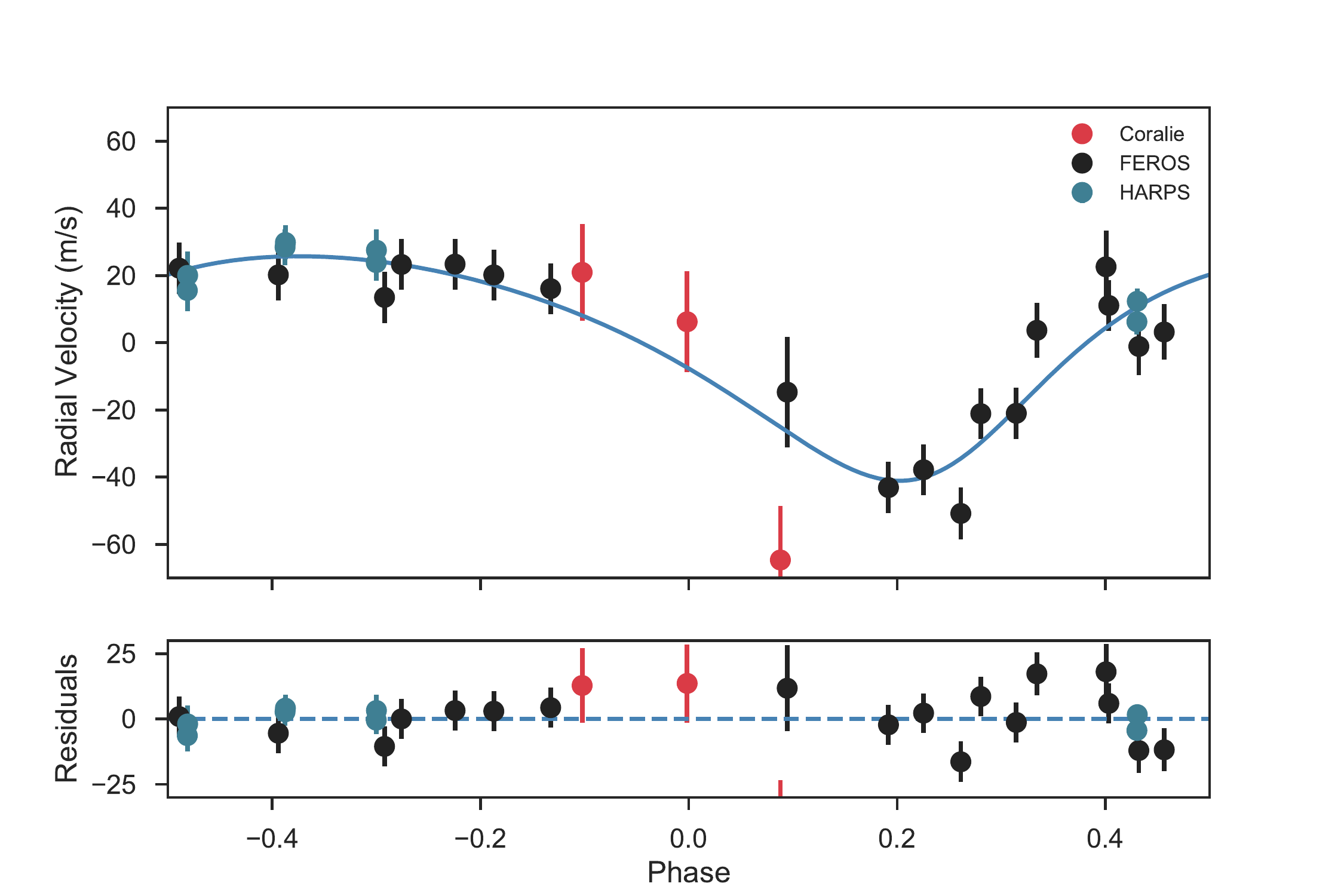}
 \caption{ The top panel presents the radial velocities (colored circles) obtained with the three spectrographs as a function of the orbital phase.
 The RV model with the derived orbital parameters for EPIC247098361b is also plotted (blue line).The bottom panel shows the residuals obtained
 for these radial velocity measurements.
 }  
 \label{exonailerrv}
\end{figure}

We performed a joint analysis of the \textit{Kepler} K2 data and follow-up radial velocities in order to determine the transit and orbital parameters of the
planetary system. For this purpose we used the \texttt{exonailer} code which is described in detail in \citet{espinoza:2016:exo}.
Briefly, we model the transit light curves using the \texttt{batman}
package  \citep{kreidberg:2015} and  we fit  them  with  the  resampling
method  described in \citet{ kipping:2013} in order to account for the smearing effect of the K2 long-cadence light curves. Following the results of \citet{EJ:2015}, we fit for the limb-darkening coefficients simultaneously with the other transit parameters, and followed \citet{espinoza:2016:lds} 
to select the  quadratic  limb-darkening  as  the  optimal
law to use for the case of EPIC247098361, as this law provides the lowest expected mean-squared error
in the planet-to-star radius ratio. The limb-darkening coefficients were fit using the uninformative sampling technique of \citet{Kipping:LDs}. A photometric jitter term was also included in the fit of the photometry, in order to empirically estimate the noise of the light curves. The radial velocities are modelled with the \texttt{rad-vel} package \citep{fulton:2018}, where we consider a
different systemic velocity and jitter factor for each instrument. Additionally, we
consider the eccentricity and argument of periastron passage as free parameters with uniform priors (an eccentric fit to the whole dataset is preferred to a circular model with a $\Delta \textnormal{BIC}=14$ in favor of the eccentric fit), and put a prior on $a/R_*$ using the value obtained by our procedures described in \S~\ref{sec:stellar-parameters}, which gave $a/R_* = 19.30 \pm 0.35$, a more precise value than the one obtainable from the transit light curve alone. The priors and posteriors of our
modelling are listed in Table~3. The transit and radial velocity models
generated from the posterior distribution are presented in Figures \ref{exonailerlc} and \ref{exonailerrv}, along with the observed data.
We used these transit and orbital parameters to obtain
the physical parameters of the planet by using the stellar properties obtained in the
previous subsections. We found that EPIC247098361b has a Saturn-like mass of 
M$_P$ = 0.397 $\pm$ 0.037 M$_J$, a Jupiter-like radius of R$_P$ = 1.000 $\pm$
0.020 R$_J$, and an equilibrium temperature of T$_{eq}$ = 991 $\pm$ 12 K.
Additionally we found that the planet has a significantly eccentric orbit of $e$ = 0.258 $\pm$ 0.025.

\begin{table*}
 \centering
 \begin{center}
   \caption{Transit, orbital, and physical parameters of EPIC247098361b. On the priors, $N(\mu,\sigma)$ stands for a normal distribution with mean $\mu$ and standard deviation $\sigma$, $U(a,b)$ stands for a uniform distribution between $a$ and $b$, and $J(a,b)$ stands for a Jeffrey's prior defined between $a$ and $b$.}
   \label{tab:planet}
 \begin{threeparttable}

  \begin{tabular}{@{}lcc@{}}
  \hline
   
  Parameter      &   Prior &  Value \\
 \hline
 Light-curve parameters \\
 
P (days) & $N(11.169,0.1)$ & 11.168454 $\pm$ 0.000023\\
T$_0$ (days) & $N(2457825.350,0.1)$ & 2457825.3497822732 $\pm$ 0.000093\\
R$_P$/R$_{\star}$ & $U(0.001,0.2)$ & 0.08868$^{+0.00044}_{-0.00042}$\\
$a/$R$_{\star}$ & $N(19.30,0.35)$ & 19.25$^{+0.27}_{-0.31}$\\
$i$ & $U(70,90)$  & 89.14$^{+0.13}_{-0.11}$ \\
q$_1$ & $U(0,1)$ & 0.417$^{+0.038}_{-0.037}$\\
q$_2$ & $U(0,1)$ & 0.318$^{+0.029}_{-0.028}$\\
$\sigma_w$ (ppm) & $J(10,5000)$ & 51.68$^{+0.68}_{-0.64}$\\

\hline
RV parameters\\

K  (m s$^{-1}$) &  $N(35,100)$  & $33.42^{+3.12}_{-3.02}$\\
$e$                    &  $U(0,1)$   & 0.258 $\pm$ 0.025 \\
$\omega$ (deg)   & $U(0,360)$ & 207 $^{+3.6}_{-3.8} $\\
$\gamma_{coralie}$ (km s$^{-1}$)   & $N(22.35,0.05)$ & 22.3394$^{+0.0087}_{-0.0092} $ \\
$\gamma_{feros}$ (km s$^{-1}$)   & $N(22.40,0.05)$ &   22.3965$^{+0.0023}_{-0.0022} $\\
$\gamma_{harps}$ (km s$^{-1}$)   & $N(22.40,0.05)$ & 22.3917$^{+0.0029}_{-0.0030} $ \\
$\sigma_{coralie}$ (km s$^{-1}$)   & $J(10^{-4},0.1)$ & 0.0011$^{+0.0013}_{-0.0010} $ \\
$\sigma_{feros}$ (km s$^{-1}$)   & $J(10^{-4},0.1)$ & 0.0029$^{+0.0042}_{-0.0025} $ \\
$\sigma_{harps}$ (km s$^{-1}$)   & $J(10^{-4},0.1)$ & 0.0008$^{+0.0024}_{-0.0006} $ \\

\hline
Derived parameters\\

M$_P$ (M$_{J}$)   &  -- & 0.397 $\pm$  0.037 \\
R$_P$ (R$_J$])      & -- & 1.000$_{-0.020}^{+0.019}$  \\
%T$_{eq}$(K)            &  -- & 992$_{-12}^{+13}$   \\
$\langle$T$_{eq} \rangle$$^{a}$ (K)	&  -- & 1030 $\pm$ 15   \\
$a$  (AU)&  -- & 0.10355$_{-0.00076}^{+0.00078}$   \\
\hline

\end{tabular}
*Time averaged equilibrium temperature using equation 16 of \citet{mendez:2017}.\\
  \end{threeparttable}
 \end{center}
\end{table*}

\subsection{Rotational modulation and search of additional transits}

A search for additional transits was performed on the photometry using the BLS algorithm \citep{BLS} with the transits of EPIC 247098361b masked out. No significant signals were found, which puts a limit of $\approx 1.5R_\oplus$ to any transiting companion orbiting in periods smaller than $\approx 38$ days. Additionally, a Generalized Lomb-Scargle periodogram \citep{zk:2009} was ran in order to search for any periodic signals in the data, but the only periods that stood out were at $1.04$ and $0.74$ days, most likely instrumental as the phased data at those periods does not show any significant, physically interpretable signal. No secondary eclipses or phase curve modulations were found in the data, which is expected given that the secondary eclipse amplitude due to reflected light would be at most $(R_p/a)^2=21 \pm 0.71$ ppm, significantly below the photometric precision of 51 ppm. 

\section{Discussion}
\label{sec:disc}

\begin{figure*}
 \includegraphics[width=\textwidth]{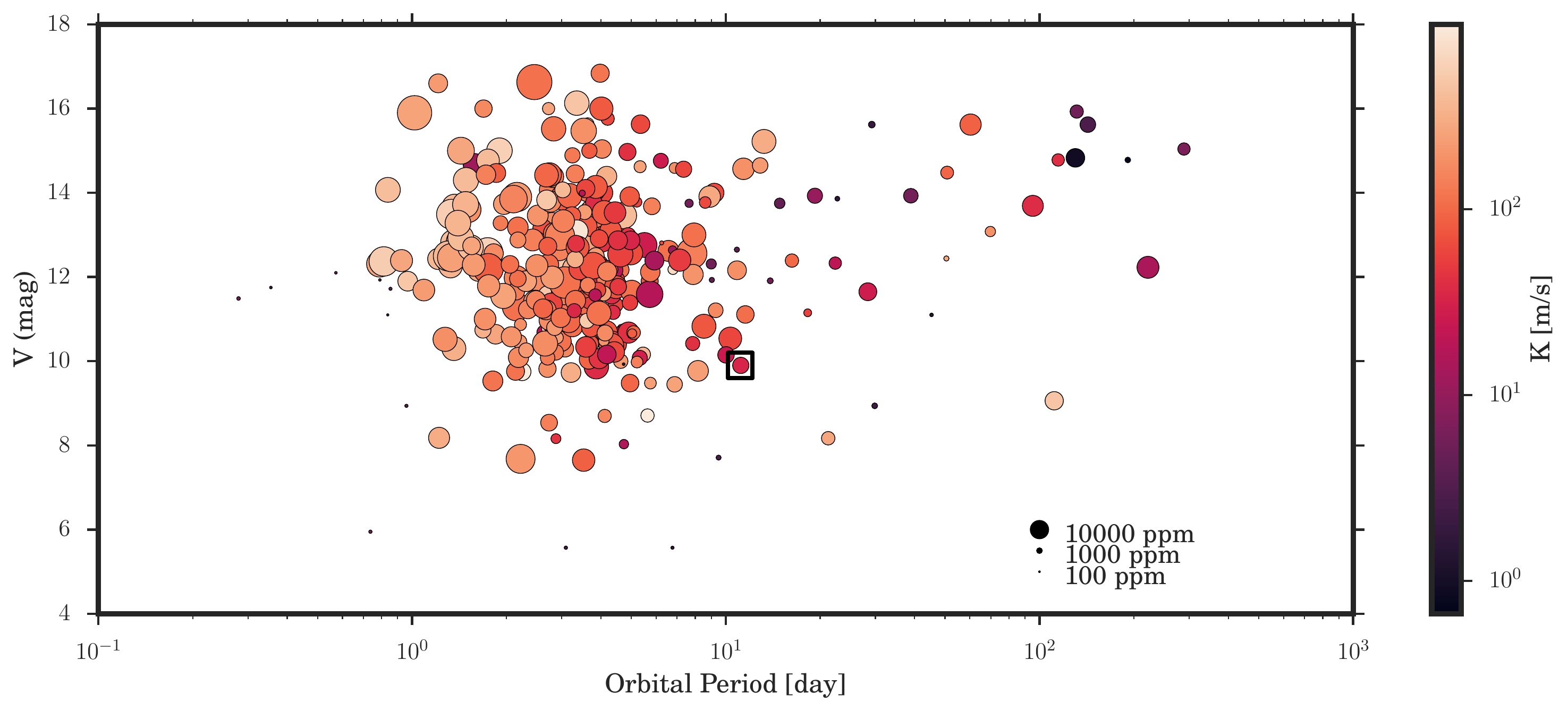}
 \caption{V magnitude as a function of orbital period for the population
 			of transiting planets with masses and radii measured with a precision
            better than 20\%. The size of the points represent the
            transit depth, while the color is related to the radial velocity
            semi-amplitude. EPIC247098361b (inside the black square) lies
            in a sparsely populated region and is one of the few giant planets with
            P$>$10d and V$<$10.}
 \label{vmag}
\end{figure*}

\begin{figure*}
 \includegraphics[width=\textwidth]{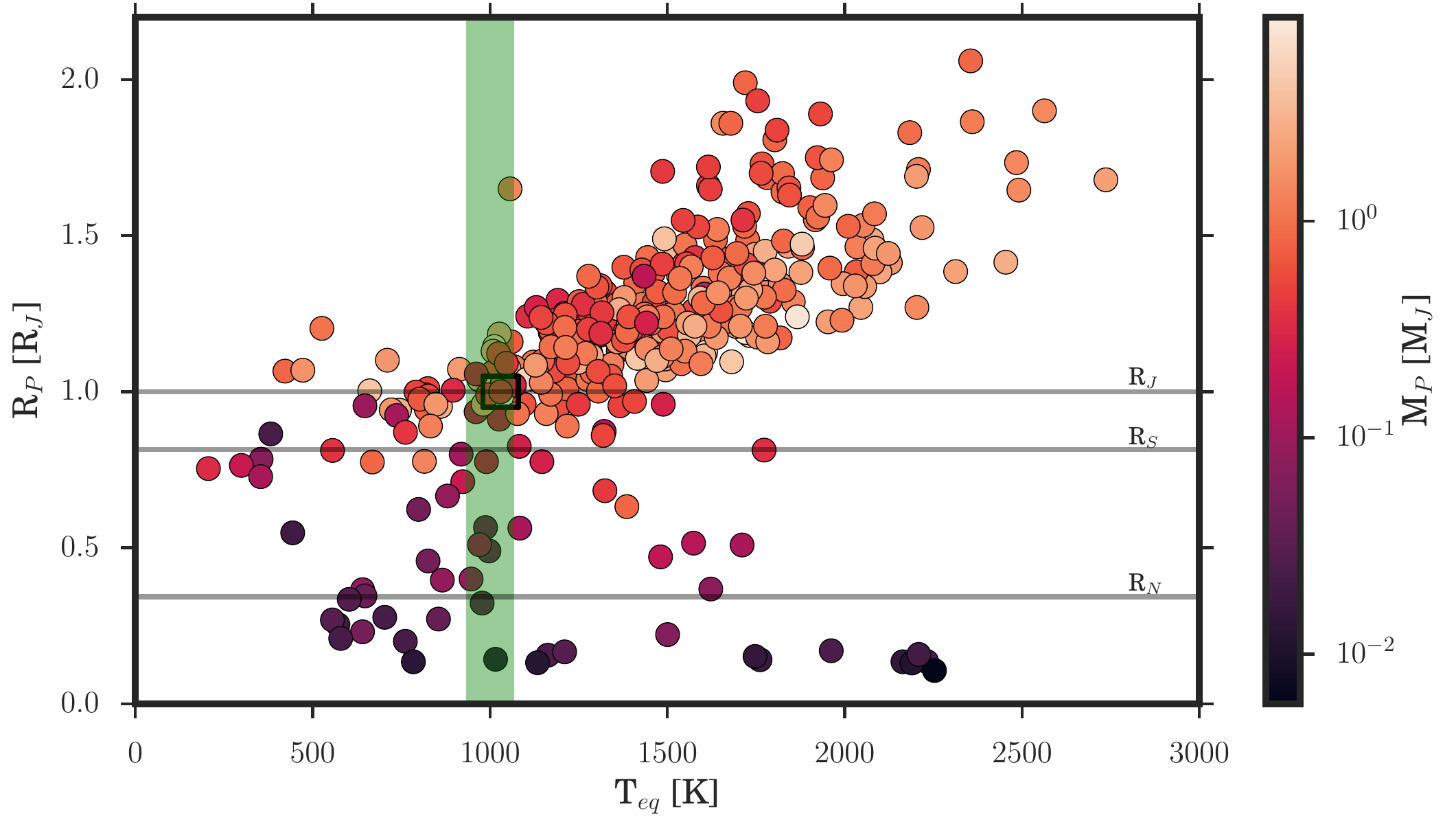}
 \caption{Population of transiting giant planets with radii and masses
 			determined at better than 20\% precision level plotted in the
        	T$_{eq}$--R$_{\star}$ plane. The green vertical bar
            indicates the transition region for which planets having
            lower equilibrium temperatures have structures that are not significantly
            affected by the inflation mechanism of hot Jupiters,
            EPIC247098361b (inside the black square), lies just inside this
            transition region.  }
 \label{teqrp}
\end{figure*}

EPIC247098361b is compared with the full population of transiting planets
with available determinations of radii and masses at the level of 20\% in Figures
\ref{vmag} and \ref{teqrp}. Due to its orbital period of $P=11.2$ d, EPIC247098361b lies in
a relatively sparsely populated region of parameter space.
Its time averaged equilibrium temperature of 1030 K lies just in the
transition region where the mechanism responsible for inflating the
radii of hot Jupiters stops playing a significant role
\citep{kovacs:2010,demory:2011}. Additionally, its current planet-to-star
separation at pericenter is large enough that the effects that
tidal and/or magnetic interactions can have on the structure and orbital
evolution of the system are expected to be small \citep{dawson:2014}.

EPIC247098361b is remarkably similar to WASP-117b \citep{lendl:2014}.
Both have have Saturn-like masses, Jupiter-like radii, eccentricities
close to 0.3, orbital periods slightly longer that 10 days, and late
F-type host stars. WASP-117b has a slightly less metal rich host star
than EPIC247098361 and its density is lower. Both systems are excellent
targets for performing detailed follow-up observations to further
understand the structure and evolution of giant planets that are
not affected by proximity effects. According to TEPCat \citep{tepcat}, there are only
$\approx 20$ other well characterised transiting giant planets with periods
longer than 10 days. EPIC247098361b ($V=9.9$) stands out as the
system with the brightest host star after HD~17156 \citep[$V=8.2$,][]{barbieri:2007}
and HD~80606 \citep[$V=9.1$,][]{moutou:2009}.

\subsection{Structure}
Due to the moderately low insolation levels received from its parent star,
the internal structure of EPIC247098361b can be studied by comparing its
measurements of mass and radius with the predictions of theoretical models.
We used the \citep{fortney:2007} models of planetary structure and evolution to
determine the mass of a possible central rocky core. These simple models assume
that all the solid material is concentrated in this core, which is likely a
simplification of the problem but serves as an illustration of the possible
internal composition of the planet. We find that, given the evolutionary
status of the EPIC247098361 system, the measured mass and radius of the planet
are consistent with having a relatively massive central core of M$_{core}$ = 20
$\pm$ 7 M$_{\oplus}$. This value is also consistent with the relation found by
\citet{thorngren:2016} between the mass of the planet and the total mass in
heavy elements. This relation was obtained using the properties of the known
transiting warm giant planets and more realistic models in which the solids are
also mixed in an H/He dominated envelope.
The prediction for the heavy element mass of  EPIC247098361b is M$_Z$=33$\pm$12 M$_{\oplus}$,
where in this case only 10 M$_{\oplus}$ of solids are located in the central core,
and a large amount of this material is required to be distributed in the planet
envelope to reproduce the observed radius of EPIC247098361b.
These properties are consistent with the core accretion model of planet formation \citep{pollack:96}
in which the planet starts a runaway accretion of gaseous material as soon as
the solid embryo reaches a mass of 10 M$_{\oplus}$. In this process,
the planet keeps accreting rocky and icy planetesimals that have been decoupled
from the gaseous disc.

EPIC247098361b is a suitable system on which to use envelope-enriched
models to get an independent estimate of M$_Z$. If put in the context
of the full population of transiting warm giant planets, this estimate
of M$_Z$ can be used to probe for correlations with other physical and
orbital properties. Specifically, \citet{miller:2011} found a tentative
correlation between the M$_Z$ and the stellar [Fe/H], which then
was put in to question by \citet{thorngren:2016} using a bigger sample
of systems and new structural models. Nonetheless, the parameters used
in the \citet{thorngren:2016} study were not obtained following an
homogeneous procedure, and additionally $\approx 25\%$ of their sample consisted
of low irradiated systems with orbital periods of $P<5$ days, whose structure
can suffer from other proximity effects (e.g. tidal, magnetic).
The combination of GAIA parallaxes, allowing an homogeneous characterization of
the host stars, coupled to new discoveries of transiting giant planets with
$P>10$ days by new ground-based (e.g. HATPI\footnote{https://hatpi.org/}),
and space-based missions \citep[e.g. TESS,][]{tess}, will be fundamental for linking the
inferred heavy element content of the planets with the global properties of
the systems.

\subsection{Migration}

While the current eccentricity of
EPIC247098361b is too low to produce significant 
migration by tidal friction \citep{jackson:2008}, it can still
be migrating if the system is being affected by secular gravitational 
interactions produced by a third distant body
\citep{kozai:62,lidov:62,li:2014,petrovich:2015}. These interactions produce 
periodic variations of the eccentricity and inclination of the system, where
the interior planet migrates during the very high eccentricity stage by tidal friction, but most of the time the planet presents
moderate eccentricities. \citet{petrovich:2016} presented  a model in which only 20\% of the warm Jupiter population
is migrating through this process. While this conclusion was reached from the eccentricity distribution of radial velocity discovered planets,
a stricter test will need a study of the distribution of spin-orbit angles of warm Jupiters. The number of warm
Jupiters with measured spin-orbit angles is still low (only 10 studied systems according to TEPcat), and EPIC247098361b is a well
suited target for measuring this angle through the Rossitter-McLaughlin effect.

\subsection{Possible follow-up observations}
The bright host star coupled to the nearly equatorial declination of the system makes of EPIC247098361
one of the most promising warm giant planets to perform detailed follow-up observations using Northern
and Southern facilities.

EPIC247098361b is a well suited system to study the atmospheres of moderately low irradiated giant planets.
While its expected transmission signal is $\delta_{trans}\approx450$ ppm, which is small compared to that of typical hot Jupiter systems, there have been reported measurements of transmission spectra for
systems with $\delta_{trans}<500$ and transit depths similar to that of EPIC247098361b. The system is specially interesting for atmospheric studies since it has been predicted that warm Saturns like EPIC247098361b, given its metal enrichment, should have low C/O ratios as compared to that of their host stars \citep{Espinoza:2017}, a picture that has been also predicted for their hotter counterparts from population synthesis models \citep{mordasini:2016} and thus this might be an excellent system to put this picture to test. In addition, the eccentricity of the system is interesting as different temperature regimes may be at play during transit and secondary eclipse, providing an interesting laboratory for exoplanet atmosphere models.

The EPIC247098361 system is also an ideal target for measuring the spin-orbit angle through the 
Rossitter-McLaughlin effect. The estimated $v\sin{i}=4.2$ km $^{-1}$ coupled to the planet-star size ratio
would produce a anomalous radial velocity signal with a semi-amplitude of $\approx$ 35 m s$^{-1}$ for an aligned orbit,
which is similar to the orbital semi-amplitude of the system, and which can be measured by numerous
spectroscopic facilities.

Warm Jupiters have been proposed to have a significant number of companions in comparison to
hot Jupiter systems \citep{huang:2016}. The presence of outer planetary-mass companions is also
required for the migration of inner planets through secular gravitational interactions \citep{dong:2014, petrovich:2016}.
EPIC247098361 can be the target of long term radial velocity follow-up observations to detect additional velocity signals
or trends that can be associated with distant companions. Finally, EPIC247098361 is a well suited
system to search for transit timing variations because its transits can be observed with relatively
small aperture telescopes from Southern and Northern facilities.

\section*{Acknowledgments}
R.B.\ gratefully acknowledges support by the
Ministry of Economy, Development, and Tourism's Millennium
Science Initiative through grant IC120009, awarded to The
Millennium Institute of Astrophysics (MAS).
 A.J.\ acknowledges support from FONDECYT project 1171208,
 BASAL CATA PFB-06, and project IC120009 ``Millennium Institute
 of Astrophysics (MAS)" of the Millennium Science Initiative,
Chilean Ministry of Economy. 
A.J.\ warmly thanks the Max-Planck-Institut f\"ur Astronomie for the hospitality during a sabbatical year where part of this work was done.
M.R.D.\ acknowledges support by CONICYT-PFCHA/Doctorado Nacional-21140646, Chile.
J.S.J.\ acknowledges support by FONDECYT project 1161218 and partial support by BASAL CATA PFB-06.
This paper includes data collected by the K2 mission. Funding
for the K2 mission is provided by the NASA Science Mission directorate.
Based on observations collected at the European Organisation for Astronomical
Research in the Southern Hemisphere under ESO programme 0100.C-0487(A).

\bibliographystyle{mnras}
\bibliography{main}

%\appendix
\begin{table*}
\label{tab:rvs}
 \centering
  \caption{Radial velocity and bisector span measurements for EPIC247098361.}
  \begin{tabular}{@{}lccccr@{}}
  \hline
     BJD              &   RV                  & $\sigma_{RV}$ &   BIS               &  $\sigma_{BIS}$ &  Instrument \\
    (-2400000)     &   [km s$^{-1}$]  & [km s$^{-1}$]   &   [km s$^{-1}$] &   [km s$^{-1}$] &   \\
 \hline

58030.8561386 & 22.4191 & 0.0103 & -0.016 & 0.012 & FEROS   \\ 
58053.8165988 & 22.3997 & 0.0077 &  0.007 & 0.010 & FEROS   \\
58058.7436015 & 22.3603 & 0.0090 &  0.001 & 0.011 & Coralie \\
58059.8696367 & 22.3456 & 0.0100 & -0.041 & 0.015 & Coralie \\
58060.8694368 & 22.2747 & 0.0116 & -0.038 & 0.017 & Coralie \\
58062.8040764 & 22.3457 & 0.0072 & -0.009 & 0.009 & FEROS   \\
58064.6920866 & 22.3979 & 0.0037 &  0.014 & 0.005 & HARPS   \\
58064.6959210 & 22.4040 & 0.0037 &  0.014 & 0.005 & HARPS   \\
58064.7125895 & 22.3954 & 0.0081 & -0.013 & 0.010 & FEROS   \\
58065.6785670 & 22.4072 & 0.0061 &  0.006 & 0.008 & HARPS   \\
58065.6823333 & 22.4117 & 0.0070 & -0.004 & 0.009 & HARPS   \\
58066.7267554 & 22.4201 & 0.0053 &  0.009 & 0.007 & HARPS   \\
58066.7305231 & 22.4215 & 0.0050 &  0.011 & 0.006 & HARPS   \\
58067.7048291 & 22.4155 & 0.0053 &  0.025 & 0.007 & HARPS   \\
58067.7063673 & 22.4192 & 0.0061 &  0.032 & 0.008 & HARPS   \\
58105.6163164 & 22.3818 & 0.0162 &  0.026 & 0.018 & FEROS   \\
58106.7021532 & 22.3534 & 0.0071 &  0.024 & 0.009 & FEROS   \\
58107.6920873 & 22.3754 & 0.0070 &  0.017 & 0.008 & FEROS   \\
58112.6484367 & 22.4198 & 0.0070 & -0.010 & 0.008 & FEROS   \\
58113.6385671 & 22.4167 & 0.0070 &  0.007 & 0.007 & FEROS   \\
58123.6355085 & 22.4100 & 0.0070 &  0.006 & 0.008 & FEROS   \\
58130.6301555 & 22.4002 & 0.0076 &  0.018 & 0.010 & FEROS   \\
58132.6018895 & 22.4187 & 0.0070 &  0.010 & 0.008 & FEROS   \\
58133.6652444 & 22.4167 & 0.0070 &  0.009 & 0.009 & FEROS   \\
58135.5601350 & 22.4199 & 0.0070 &  0.003 & 0.009 & FEROS   \\
58136.5854351 & 22.4126 & 0.0070 & -0.001 & 0.008 & FEROS   \\
58140.5838892 & 22.3587 & 0.0070 &  0.002 & 0.008 & FEROS   \\
58141.5782837 & 22.3755 & 0.0070 &  0.005 & 0.008 & FEROS   \\
58142.5699832 & 22.4076 & 0.0070 &  0.010 & 0.008 & FEROS   \\

\hline

\end{tabular}
\end{table*}
%\section[]{}

\label{lastpage}

\end{document}